\documentclass{article}

\usepackage{arxiv}

\usepackage[utf8]{inputenc} 
\usepackage[T1]{fontenc}    
\usepackage{url}            
\usepackage{booktabs}       
\usepackage{amsfonts}       
\usepackage{nicefrac}       
\usepackage{microtype}      
\usepackage{lipsum}		
\usepackage{graphicx}
\usepackage{natbib}
\usepackage{doi}

\usepackage{graphicx}
\usepackage{multicol,multirow}
\usepackage{amsmath,amssymb,amsfonts}
\usepackage{mathrsfs}
\usepackage{mathalpha}
\usepackage{amsthm}
\usepackage{rotating}
\usepackage{appendix}
\usepackage{ifpdf}
\usepackage[T1]{fontenc}
\usepackage{newtxtext}
\usepackage{newtxmath}
\usepackage{textcomp}
\usepackage{xcolor}
\usepackage{lipsum}
\usepackage{hyperref}

\title{Z-Curve Plot:\\A Visual Diagnostic for Publication Bias in Meta-Analysis}

\author{ \href{https://orcid.org/0000-0002-0018-5573}{\includegraphics[scale=0.06]{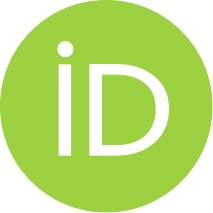}\hspace{1mm}František Bartoš} \\
	Department of Psychological Methods\\
	University of Amsterdam\\
	\texttt{f.bartos96@gmail.com} \\
	\And
	\href{https://orcid.org/0000-0001-9456-5536}{\includegraphics[scale=0.06]{orcid.pdf}\hspace{1mm}Ulrich Schimmack} \\
	Department of Psychology\\
	University of Toronto, Mississauga\\
	\texttt{ulrich.schimmack@utoronto.ca} \\
}

\hypersetup{
pdftitle    = {Z-Curve Plot: A Visual Diagnostic for Publication Bias in Meta-Analysis},
pdfsubject  = {stat.ME},
pdfauthor   = {František Bartoš, Ulrich Schimmack},
pdfkeywords = {Meta-Analysis, Publication Bias, Z-Curve, Model Fit},
}

\begin{document}
\maketitle

\begin{abstract}
Publication bias undermines meta-analytic inference, yet visual diagnostics for detecting and understanding model misfit due to publication bias are lacking.
We propose the z-plot, a visual publication bias-focused absolute model fit diagnostic.
The z-plot overlays the model-implied distribution of $z$-statistics on the observed distribution of $z$-statistics.
Models that approximate the data well show minimal discrepancy between the observed and predicted distributions of $z$-statistics, whereas models that approximate the data poorly show systematic discrepancies.
Discontinuities in the observed distribution of $z$-statistics at significance thresholds or at zero provide visual evidence of publication bias; models that account for this bias track these discontinuities.
In addition, the z-plot facilitates visual model fit comparison of competing meta-analytic models within a single figure.
We demonstrate the visualization and its interpretation with a Bayesian random-effects meta-analysis, a Bayesian PET model, a Bayesian three-parameter selection model, and RoBMA on simulated datasets and a real meta-analysis.
The method is implemented in the \texttt{RoBMA} \texttt{R} package.
\end{abstract}


\section{Introduction}

An important concern for the validity of meta-analytic conclusions is publication bias, the suppression of statistically non-significant and nonconforming findings \citep{sterling1959publication, rosenthal1964further}.
This well-documented suppression \citep{scheel2021excess, zwet2021significance, masicampo2012peculiar} causes overestimated effect sizes and inflated evidence against the null hypothesis \citep{de2015surge, fanelli2017meta, bartos2022footprint}.
As such, numerous methods were developed to adjust for publication bias, including selection models \citep{hedges1984estimation, iyengar1988selection, vevea1995general, copas1999works, van2015meta, citkowicz2017parsimonious}, regression (asymmetry) based approaches \citep{duval2000trim, stanley2014meta, bom2019kinked}, and Bayesian model-averaging methods \citep{maier2020robust, bartos2022robust} (also see \citep{irsova2023spurious} for selection on standard errors and \citep{marks2020historical} for a historical overview).

Despite these methodological developments, there has been a lack of publication bias-specific visual diagnostics.
An exception is the well-known funnel plot \citep{light1984summing} with several extensions, such as contour enhancement \citep{peters2008contour} and power enhancement \citep{kossmeier2020power}.
However, funnel plots are criticized due to their unreliability in detecting publication bias \citep{terrin2005empirical, lau2006case} and their vulnerability to ``small-study effects''.\footnote{
In other words, the funnel plot might falsely indicate publication bias in settings where studies of different sizes target different populations (\citep{tang2000misleading}; e.g., small clinical trials vs. large general-population trials), and other settings \citep[see][for an overview]{egger1997bias}. While some of the limitations can be partially addressed by incorporating covariates into the analysis, any detected asymmetry can reflect genuine heterogeneity or other small-study effects rather than publication bias \citep{sterne2011recommendations}, which invites the dismissal of publication bias concerns.}
More importantly, while funnel plots are valuable tools for data exploration before statistical modeling takes place, they are not meant for the visual assessment of already estimated models.

We propose a publication bias-specific meta-analytic model fit visualization inspired by z-curve \citep{brunner2020estimating, bartovs2020z} called ``z-plot'' that can accompany statistical tests of model fit.
The visualization allows researchers to a) assess the absolute fit of meta-analytic models to data (answering the question ``Does any of the methods describe the data well?''; \citep{box1976science}) and b) simultaneously compare the fit of different meta-analytic models to data (answering the question ``Which method describes the data better?'').
The z-plot overlays the model-implied distribution of $z$-statistics on the observed distribution of $z$-statistics.
Publication bias is indicated by sharp discontinuities in the observed data---typically at the significance threshold (selection for statistical significance) or at zero (selection for positive results).
Meta-analytic models that ignore these discontinuities misfit the data and should not be used for inference; models that respect them provide a better basis for inference.
Conversely, the absence of such discontinuities and the close agreement between model-implied and observed distributions increase confidence in unadjusted models.

In the following sections, we describe the methodology and apply it to a real example.
A vignette (\url{https://cran.r-project.org/web/packages/RoBMA/vignettes/v36-zplot.html}) provides code with additional examples and Appendix~A derives additional publication bias metrics quantifying the discrepancy between the observed and expected data.
The methodology is implemented in the \texttt{RoBMA} \texttt{R} package \citep{RoBMA}.

\section{Methodology}
\label{sec:methods}

We illustrate the methodology using two simulated datasets.
The first ``unbiased'' dataset contains $300$ effect size estimates from two-sample t-tests simulated in the absence of publication bias, with average effect Cohen's $d = 0.30$, between-study heterogeneity $\tau = 0.15$, and per-group sample sizes drawn from a truncated inverse-gamma distribution, $n \sim $ Inverse-Gamma$_{[5, 1905]}$(shape $=1.153$, scale $= 0.046$).\footnote{
This simulation setting is based on the \citet{carter2019correcting} simulation study implemented in the \texttt{PublicationBiasBenchmark} \texttt{R} package.}
The second ``biased'' dataset is created by applying publication bias to the unbiased dataset.
We retain statistically non-significant or negative studies with a probability of $50$\%, which results in retaining $182$ estimates.
As such, the first dataset is generated by a random-effects model and the second by a three-parameter selection model (3PSM).

\begin{figure}
    \centering
    \includegraphics[width=0.9\linewidth]{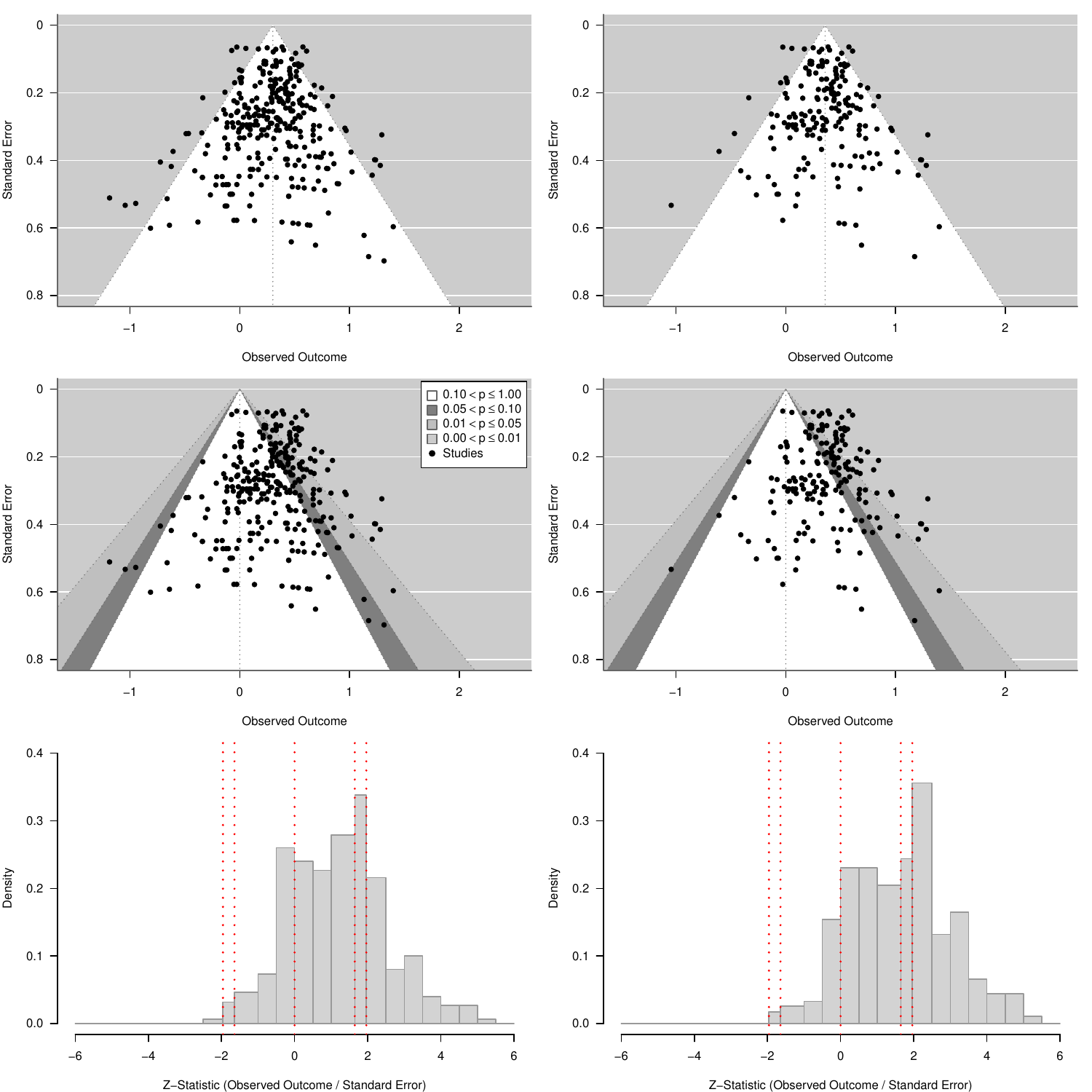}
    \caption{Standard Funnel Plot, Contour-Enhanced Funnel Plot, and Selection-Enhanced z-histogram of Simulated Datasets}
    \begin{flushleft}
    \small
    \textit{Note.} The left panels visualize the $300$ simulated studies in the absence of publication bias.
    The right panels visualize the 182 studies that survive the publication process after applying publication bias to the same data.
    The first row consists of the standard funnel plots centered at the random-effects estimate, the second row consists of the contour enhanced funnel plots centered at zero, and the third row consists of histograms of $z$-statistics with bins and vertical lines highlighting the usual selection thresholds.
    \end{flushleft}
    \label{fig:funnels}
\end{figure}

Figure~\ref{fig:funnels} visualizes the unbiased (left column) and biased (right column) datasets using funnel plots (first row) and contour enhanced funnel plots (generated using the \texttt{metafor} \texttt{R} package; \cite{metafor}).
The figure highlights two limitations of funnel plots.
First, despite the $118$ missing estimates, the funnel plot of the biased dataset does not show clear funnel plot asymmetry and does not reveal the existing publication bias, Egger's test $t(180) = 1.26$, $p = 0.210$.\footnote{
Even if a funnel plot asymmetry were present, it could have resulted from small study effects, i.e., smaller studies may legitimately target larger effects, resulting in the same asymmetric pattern in the absence of publication bias \citep{tang2000misleading, sterne2011recommendations}.}
Second, the funnel plot of the biased dataset does not indicate that a publication bias-unadjusted model is ill-suited for the data and does not facilitate visual comparison of different model fits.
Both of these limitations partially stem from the two-dimensional nature of the funnel plot.
In the first case, the pattern of publication bias is spread across a two-dimensional area, which dilutes the signal.
In the second case, evaluating model fit to a multi-dimensional representation of data (or comparing multiple model fits to it) is genuinely difficult.

\subsection{Distribution of observed $z$-statistics}

One way of addressing those limitations is examining a simpler projection of the data such as $z$-statistics (observed effect sizes $\text{y}$ divided by their standard errors $\text{se}$).
$Z$-statistics are an interesting representation of the data because they map to the usual target of publication bias, $p$-values, while retaining high resolution in the dense area of the data (i.e., mostly small $p$-values, which inadvertently become ``squished together'' when examining the distribution of $p$-values directly).
An important feature of the sampling distribution of $z$-statistics---under the absence of publication bias---is that it is a smooth mixture of approximately normal distributions.\footnote{
Although some outcomes yield discrete distributions of test statistics \citep[e.g., binary data;][]{viechtbauer2021model}, we would not expect sharp discontinuities when aggregating across trials of various sample sizes.}
Discontinuities at statistical significance thresholds ($|z|=1.96$, $|z|=1.64$) or direction of the effect ($z=0$) are directly indicative of publication bias.
Furthermore, such discontinuities, unlike funnel plot asymmetry, are unlikely under small-study effects and cannot be attributed to hidden moderators, which could, in the extreme case, cause a \emph{smooth} multimodal distribution.

The third row of Figure~\ref{fig:funnels} shows selection-enhanced z-histograms of the simulated datasets: red dotted vertical lines highlight the statistical significance thresholds, and the binning algorithm incorporates the thresholds.
The selection-enhanced z-histogram of the unbiased dataset (left column) does not show any stark discontinuities, and as such, it does not indicate publication bias.
However, the selection-enhanced z-histogram of the biased dataset (right column) shows a suspicious discontinuity at the statistical significance threshold ($z = 1.96$), which was difficult to detect from the funnel plots.

In fact, histograms of large collections of $z$-statistics similar to those in the third row of Figure~\ref{fig:funnels} have become an intuitive display of severe publication bias in recent years.\footnote{
See the ``Dark Statistics'' meme by Kareem Carr: \url{https://x.com/kareem_carr/status/1335282974443655168}.}
In addition, several methods such as z-curve (\cite{brunner2020estimating}; see also \cite{brodeur2016star, zwet2021significance}) were developed to examine the distribution of $z$-statistics from large sets of studies and assess the quality of the underlying literature \citep[see e.g.,][also for additional reasons justifying the use of $z$-statistics rather than $p$-values directly]{schimmack2023estimating, gupta2023tempest}.
Those methods differ from the z-plot described below since they are directly concerned with the distribution of test statistics, typically estimating it via mixtures of (truncated) normal distributions, and do not provide standard meta-analytic inference.

\subsection{Model implied distribution of $z$-statistics}

The selection-enhanced z-histogram can be extended with the implied sampling distribution of the $z$-statistics under the fitted meta-analytic models.
Comparison of the observed distribution of the $z$-statistics with the model-implied distribution of $z$-statistics allows us to assess whether the estimated model fits the observed data well.
If the model fits the data well, the model-implied distribution of $z$-statistics should closely follow the histogram of observed $z$-statistics.
While some histogram bins are expected to be higher or lower than the model-implied distribution due to sampling variance under a good model fit (especially when the number of estimates is small), systematically and notably higher or lower bins on opposite sides of selection thresholds identify model misfit due to publication bias.
Additionally, we can visually compare the fit of the implied distribution of the $z$-statistics under several competing models to gain insights into the (mis)fit of the different models.

In Bayesian settings, the implied distribution of the $z$-statistics based on a model $\pi$ with parameters $\theta$ corresponds to the posterior predictive distribution of $z$-statistics $\text{z}_\text{pred}$ \citep[see][for a detailed introduction to posterior predictive distributions]{gabry2019visualization, gelman2020bayesian, schad2022workflow} and can be computed in the following way.
First, we obtain the posterior predictive distribution for the effect sizes $\text{y}_\text{pred}$,
\begin{equation*}
    \pi(\text{y}_\text{pred} \mid y, \text{se}) = \int \pi(\text{y}_\text{pred} \mid \theta, \text{se}) \, \pi(\theta \mid \text{y}, \text{se}) \, \mathrm{d}\theta.
\end{equation*}
Then, we perform a change of variables $\text{z}_\text{pred} = \text{y}_\text{pred}/\text{se}$ to obtain the posterior predictive distribution for $z$-statistics,
\begin{equation*}
    \pi(\text{z}_\text{pred} \mid y, \text{se}) = \int \text{se} \, \pi\!\left(\text{se} \, \text{z}_\text{pred} \mid \theta, \text{se}\right) \, \pi(\theta \mid \text{y}, \text{se}) \, \mathrm{d}\theta.
\end{equation*}
Overlaying this model-implied distribution on the selection-enhanced z-histogram yields the $z$-plot, which turns the comparison into a visual assessment of absolute model fit.

\begin{figure}
    \centering
    \includegraphics[width=1\linewidth]{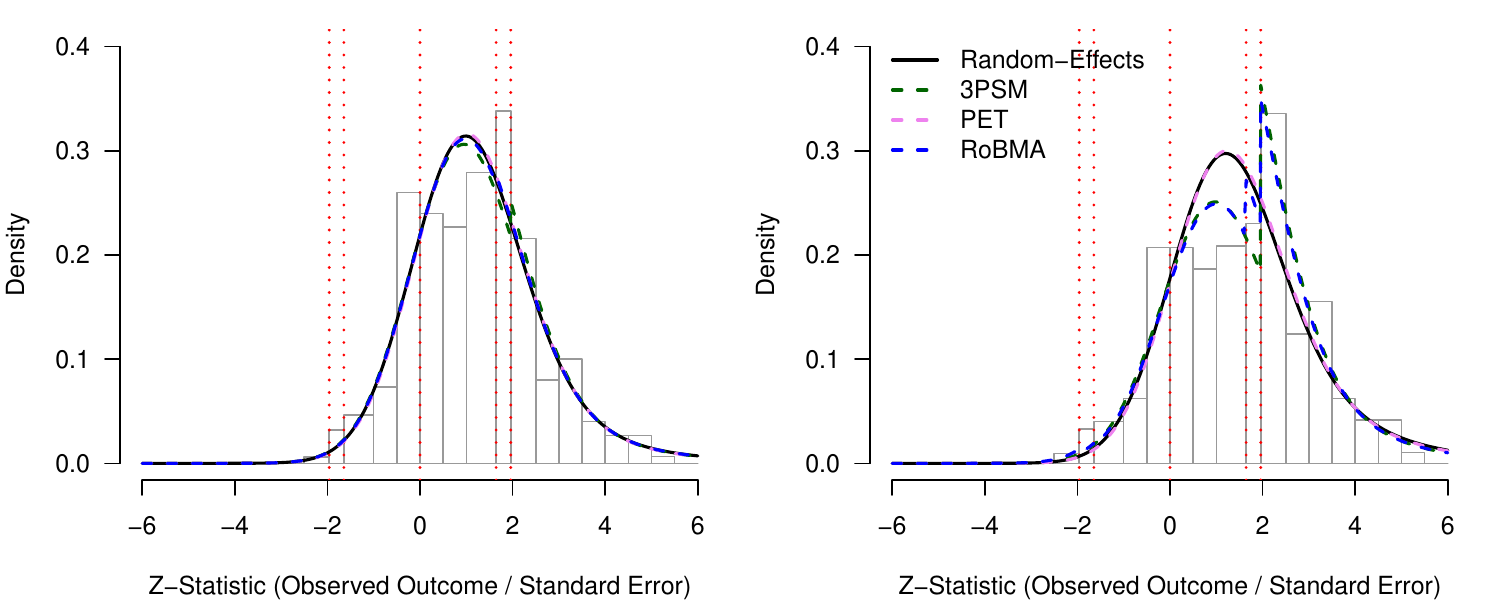}
    \caption{Z-Plot: Model Fit of the Four Meta-Analytic Models to the Simulated Datasets}
    \begin{flushleft}
    \small
    \textit{Note.} Gray bars show the observed distribution of $z$-statistics and the curves show the model-implied distributions. Dotted red lines mark $z=0$ and the significance thresholds ($z=\pm1.96$). The left panel visualizes the $300$ simulated studies in the absence of publication bias; the right panel visualizes the $182$ studies that survive under publication bias.
    \end{flushleft}
    \label{fig:zplot}
\end{figure}

Figure~\ref{fig:zplot} displays posterior predictive distributions of four different models fitted to the simulated datasets: a Bayesian random-effects model (solid black), a Bayesian three-parameter selection model (3PSM; dashed dark green), a Bayesian precision-effect test model (PET; dashed violet), and robust Bayesian meta-analysis (RoBMA; dashed blue).
The random-effects, PET, and 3PSM models closely correspond to their frequentist counterparts, whereas RoBMA applies Bayesian model averaging \citep[e.g.,][]{hoeting1999bayesian, fragoso2018bayesian, hinne2019conceptual} across meta-analytic models that make different assumptions about the presence of an effect, heterogeneity, and publication bias \citep[including six weight-functions, PET, and PEESE model; see Table 1 in][]{bartos2022robust}.
All models were estimated using the \texttt{RoBMA} \texttt{R} package \citep{RoBMA} with default prior distributions for standardized mean differences: Normal(mean $=0$, sd $=0.707$) for $\mu$ and Normal$_+$(mean $=0$, sd $=0.354$) for $\tau$; methodological details are provided in \citet{bartos2022robust}.

When publication bias is absent (left column of Figure~\ref{fig:zplot}), all models approximate the observed $z$-distribution closely.
There are only small non-systematic departures of the model-implied distribution of $z$-statistics under all models from the observed distribution of $z$-statistics due to the sampling variance of the data.
Consequently, each of the specified models can be visually assessed as an equally suitable approximation of the data with no large absolute model fit departures.
All specified models approximate the unbiased dataset well because the posterior distributions of the publication bias related parameters of the Bayesian PET, 3PSM, and RoBMA models converge to the edge of the parameter space that corresponds to a random-effects meta-analysis.\footnote{
The regression estimate from PET is estimated as $0.077$, 95\% CrI $[0.002, 0.249]$ (the prior distribution is truncated at 0), and the relative publication probability of non-significant studies is estimated as $0.871$, 95\% CrI $[0.655, 0.995]$ (the prior distribution is bounded at 1).}
The second column of Table~\ref{tab:sim-estimates} shows, in this specific case, that the mean effect size estimates from the different models closely agree.
Despite the agreement in effect size estimates, a statistical comparison of the specified models incorporates model complexity into the relative model fit assessment and highlights that the publication bias unadjusted model predicts the data several times better than any of the publication bias-adjusted models (third column of Table~\ref{tab:sim-estimates}).
As such, the visual assessment allows us to visually validate the statistical tests and interpret the random-effects meta-analysis with more confidence.

\begin{table}
    \centering
    \begin{tabular}{lrr|rr}
    Method & \multicolumn{2}{c}{Unbiased Dataset} & \multicolumn{2}{c}{Biased Dataset} \\
           & Pooled Estimate [95\% CrI] & $\text{BF}_\text{bias}$ & Pooled Estimate [95\% CrI] & $\text{BF}_\text{bias}$ \\
    \hline
    Random-Effects & $0.301$ $[0.269, 0.334]$ & $-$       &  $0.358$ $[0.320, 0.396]$ & $-$     \\
    3PSM           & $0.290$ $[0.252, 0.327]$ & $0.215$   &  $0.308$ $[0.251, 0.361]$ & $23.34$ \\
    PET            & $0.285$ $[0.236, 0.325]$ & $0.061$   &  $0.317$ $[0.247, 0.375]$ & $0.302$ \\
    RoBMA          & $0.298$ $[0.255, 0.333]$ & $0.168$   &  $0.309$ $[0.235, 0.377]$ & $6.48$  \\
    \end{tabular}
    \caption{Summary of the Four Meta-Analytic Models Fitted to the Simulated Datasets}
    \begin{flushleft}
    \small
    \textit{Note.} Posterior pooled effect size estimates with 95\% central credible intervals. The publication bias Bayes factors are computed as comparison with random-effects model for 3PSM and PET model and comparison with the sub-ensemble of publication bias-unadjusted models for RoBMA.
    \end{flushleft}
    \label{tab:sim-estimates}
\end{table}

When publication bias is present (right panel of Figure~\ref{fig:zplot}), the random-effects and PET models miss the discontinuity at the significance threshold: both overestimate the proportion of non-significant results (to the left of the threshold) and underestimate the proportion of significant results (to the right of the threshold).
In contrast, 3PSM and RoBMA approximate the discontinuity at statistical significance well and approximate the observed distribution of $z$-statistics more closely.
3PSM approximates the biased dataset well because the model directly corresponds to the defined data-generating mechanism; RoBMA approximates the biased dataset well because it correctly places the majority of posterior model probability on selection models including selection on statistical significance.
The fourth column of Table~\ref{tab:sim-estimates} shows, in this specific case, that the mean effect size estimates from the publication bias adjusted models (especially 3PSM) are much closer to the true effect size of $0.30$ than the random-effects estimate.
A statistical comparison of the specified models further highlights that especially 3PSM and RoBMA predict the data several times better than the publication bias-unadjusted model (fifth column of Table~\ref{tab:sim-estimates}).
Again, the visual assessment allows us to visually support the statistical tests and explain why the random-effects meta-analysis is not an appropriate model for the biased dataset and a publication bias-adjusted model such as 3PSM is needed.

\section{Example}

To illustrate how researchers can apply z-plot in a simplified analytic workflow, we use an example from a meta-analysis of social comparison as a behavior change technique \citep{hoppen2025socialcomparison}.
Social comparison as a behavior change technique (SC-BCT) is based on our tendency to compare ourselves against others \citep{wood1996social}.
SC-BCT provides a social standard, prompts individuals to assess their (dis)similarity to that standard, and thereby motivates them to adapt to or maintain their behavior.
In practice, SC-BCT connects individual feedback with social benchmarks, often delivered repeatedly via email, leaderboards, or letters to prompt change.
\citet{hoppen2025socialcomparison} conducted a meta-analysis of randomized controlled trials evaluating the efficacy of SC-BCT for outcomes in climate-change mitigation, health, performance, and service domains.
We examine the main comparison of SC-BCT to passive controls.
The original pooled effect was $g=0.17$, 95\% CI $[0.11, 0.23]$, based on a random-effects model examining 37 trials.
Egger's test performed by the authors was not statistically significant, $t(35) = -0.47$, $p = 0.640$, and the authors deemed publication bias unlikely.

\begin{figure}
    \centering
    \includegraphics[width=0.9\linewidth]{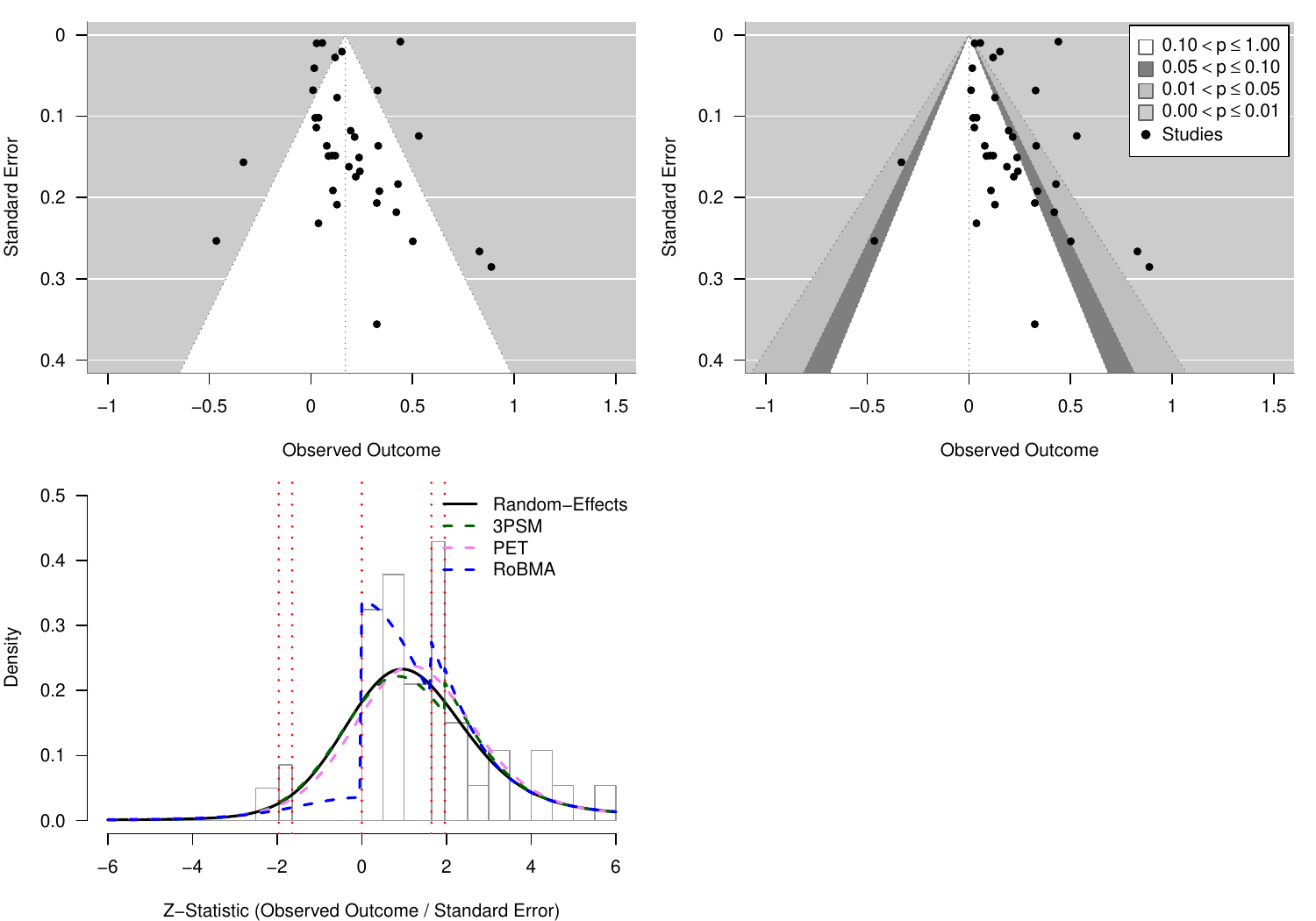}
    \caption{Funnel Plot, Contour Enhanced Funnel Plot, and z-plot Showing Strong Publication Bias in the Meta-Analysis of Social Comparison as a Behavior Change Technique}
    \label{fig:example1}
\end{figure}

In this simplified analytic workflow, let us assume that the analysts consider applying the Bayesian random-effects model, the Bayesian PET model, the Bayesian 3PSM model, and the RoBMA model as described in the previous section.
The first row of Figure~\ref{fig:example1} visualizes the traditional and contour enhanced funnel plots indicating a slight skew of the observed effect size estimates; however, as indicated by the originally reported Egger test, we cannot reject the null hypothesis of no publication bias.
Estimating and comparing the simple statistical models reveals that the 3PSM model predicts the data $3.34$ times worse than the random-effects model and the PET model predicts the data $1.32$ times better than the random-effects model (PET model predicts the data $4.40$ times better than the 3PSM model).
However, RoBMA finds that the publication bias-adjusted models that are part of its model-ensemble predict the data $440.18$ times better than publication-bias unadjusted models.
Examination of the z-plot in the second row of Figure~\ref{fig:example1} shows why: none of the simple models accounts for the selection for positive results that is apparent in the z-plot (random-effects, 3PSM, and PET systematically overestimate the number of negative estimates and underestimate the number of positive estimates).
RoBMA correctly accounts for the selection for positive results via the included selection models, which contain the corresponding step.
RoBMA also attempts to account for the inflation of marginally significant results; however, the bin still visibly exceeds the model-implied distribution.
This would suggest that a more complex model, possibly incorporating $p$-hacking into the marginal significance region, is required.
In the absence of such a model, the best possible inference for this dataset can be obtained by RoBMA, which finds moderate evidence for the absence of the effect, $\text{BF}_{10} = 0.166$, with a model-averaged effect size estimate of $-0.008$, 95\% CrI $[-0.174, 0.059]$ and between-study heterogeneity $\tau = 0.209$, 95\% CrI $[0.139, 0.307]$.

In summary, the visualization also explains why neither PET nor 3PSM adjusts for publication bias in this example---the publication bias primarily operates on the direction of the effect, which is not modeled by either model.
Either a 4PSM model or a publication bias-adjusted model ensemble such as RoBMA that contains more complex weight functions is needed to approximate the data well.
This finding adds to existing concerns about publication bias in the behavior-change literature \citep[see also][]{maier2022adjusting, bakdash2022left, szaszi2022no, beermann2024effective}.

\section{Discussion}

Z-plots provide a simple visual assessment of meta-analytic model fit aimed at detecting and understanding misfit due to publication bias.
We implemented the methodology in the \texttt{RoBMA} \texttt{R} package for all types of Bayesian meta-analytic models (i.e., meta-regression, multilevel models, and various types of publication bias adjustments) and illustrated how to interpret the z-plots using simulated and empirical examples.

One of the z-plot's benefits is that it allows simultaneous visualization of several model fits to the same observed data.
In this way, z-plot complements (relative) model-fit hypothesis tests and assessments (e.g., likelihood-ratio tests, AIC, BIC, and Bayes factors) and provides a sense of absolute fit.
The visual model assessment allows practitioners to ``see'' when simple publication bias-unadjusted models are inappropriate and gain confidence in interpreting complex publication bias-adjusted models.

The z-plots might also indicate that none of the considered models describes the data well (e.g., no model accounts for notable features of the observed distribution of test statistics).
In those cases, the z-plot might help researchers identify additional ways in which the meta-analytic model needs to be refined (or even suggest abstaining from interpreting the results altogether).
Importantly, researchers refining the model based on the observed data need to be aware that such a refinement invalidates subsequent hypothesis tests, since the same data were used twice.

We implemented the z-plots in the Bayesian framework by transforming the posterior predictive distribution.
However, similar visualizations could also be implemented within the classical framework by displaying the sampling distribution of the $z$-statistics implied by the model.
Such development would be highly beneficial to the research community, as the majority of researchers still use classical statistical techniques.

Z-plots are most informative for moderately sized or large meta-analyses.
With only a small number of estimates (e.g., 10-30 effect-size estimates), histograms of test statistics are most likely not very informative (unless all test statistics cluster exactly at the significance criterion).
The visualization relies on $z$-statistics computed as the effect size estimates divided by their standard errors.
The $z$-statistics most likely do not directly correspond to the actual test statistics performed in the original studies (e.g., $t$-statistics, $F$-statistics), which might introduce a slight distortion of the presented distribution.
However, most meta-analytic approaches (including selection models) assume normally distributed effect size estimates with known standard errors, and we would not expect meaningful changes in the majority of settings.
In our applications, we modified the default \texttt{R} histogram to increase granularity and align bin boundaries with selection thresholds (i.e., $\pm 1.96$ for statistical significance, $\pm 1.64$ for marginal significance, and $0$ for selection on the expected direction).
As with any visualization, these choices introduce some subjectivity; nevertheless, we believe that the figures convey useful information about the model fit.

Publication bias may co-occur with questionable research practices (QRPs) \citep{john2012measuring, stefan2022big, nagy2025bestiary, mathur2024phacking}.
Some QRPs are effectively equivalent to publication bias (e.g., selective reporting); however, others, such as aggressive optional stopping, might create more severe deviations in the distribution of test statistics than publication bias.
While this pattern will be clearly detectable on the z-plot, it is probably not distinguishable from the similar patterns introduced by publication bias.
Unfortunately, publication bias adjustment methods are not usually designed to handle those extreme cases of $p$-hacking \citep[e.g.,][]{mathur2024phacking} and often result in underestimating the pooled effect \citep{bartos2022robust}.
Although we did not examine QRPs in this manuscript, it might be possible that the z-plots help identify cases where publication bias adjustment models do not fit the observed data well due to QRPs.

In summary, z-plots provide a publication bias-focused visual fit diagnostic that integrates with standard meta-analytic workflows.
The visualizations complement model comparison tests and provide insight into absolute model fit.


\paragraph{Funding Statement}
None.

\paragraph{Competing Interests}
None.

\paragraph{Generative AI}
ChatGPT and Claude was used for sentence-level editing and proofreading.

\paragraph{Data Availability Statement}
The code and data are available in an annotated vignette at \url{https://osf.io/pm45q/files/osfstorage}. The \texttt{RoBMA} \texttt{R} package is available at \url{https://CRAN.R-project.org/package=RoBMA}.

\paragraph{Ethical Standards}
The research meets all ethical guidelines, including adherence to the legal requirements of the study country.

\paragraph{Author Contributions}
Conceptualization: F.B.; U.S. Methodology: F.B.; U.S. Data curation: F.B. Data visualization: F.B. Writing original draft: F.B.; All authors approved the final version of the manuscript.

\clearpage
\newpage

\setcounter{table}{0}
\setcounter{figure}{0}
\setcounter{section}{0}
\renewcommand{\thetable}{A\arabic{table}}
\renewcommand{\thefigure}{A\arabic{figure}}
\renewcommand{\thesection}{Appendix A}

\section{Additional Publication Bias Metrics}

Under adjusted models, such as PET, PEESE, selection models, or their model-averaged ensemble like RoBMA, one can extrapolate from the fitted model to the pre-publication bias state of the literature. We denote the posterior predictive distribution for the original set of studies (prior to publication bias) by $\pi^*(\text{y}_\text{pred} \mid \text{y}, \text{se})$ (extrapolated posterior predictive distribution).

For PET and PEESE, the extrapolated posterior predictive distribution is obtained by setting the standard error or sampling variance coefficients to zero (i.e., adjusting the sampling distribution for the bias associated with smaller studies). For selection models with weighted-normal likelihood
\begin{equation*}
    \text{Weighted-normal}(\text{y}|\mu, \tau^2 + \text{se}^2, \omega) = \text{Normal}(\text{y}|\mu, \tau^2 + \text{se}^2) \, w(\omega, \text{y}/\text{se}) \, I(\mu,\tau,\omega,\text{se})^{-1},
\end{equation*}
where a weight function $w$ assigns relative publication probabilities $\omega$ based on the $z$-statistic $\text{y}/\text{se}$, and the normalizing constant $I$ corresponds to
\begin{equation*}
I(\mu,\tau,\omega,\text{se}) = \int \text{Normal}(x|\mu, \tau^2 + \text{se}^2) \, w(\omega, x/\text{se}) \, \mathrm{d}x,
\end{equation*}
the extrapolated posterior predictive distribution is given by
\begin{equation*}
\pi^*_{\text{selection model}}(\text{y}_\text{pred} \mid \text{y}, \text{se}) = \text{Normal}(\text{y}_\text{pred}|\mu, \tau^2 + \text{se}^2) \, I(\mu,\tau,\omega,\text{se})^{-1}.
\end{equation*}
The multiplication by the inverse of the normalizing constant of the weighted distribution, $I^{-1} \geq 1$, is required to expand the predictive distribution proportionately to the censoring captured by the weight function. Importantly, $\pi^*_{\text{selection model}}$ is no longer a probability distribution as it integrates to $I^{-1}$ rather than to 1 with respect to $\text{y}_\text{pred}$. Those extrapolated posterior predictive distributions for the effect size $\pi^*$ can be, again, transformed to the corresponding extrapolated posterior predictive distributions for the $z$-statistics $\text{z}^*_\text{pred}$ via a change of variables.

\begin{figure}[h]
    \centering
    \includegraphics[width=1\linewidth]{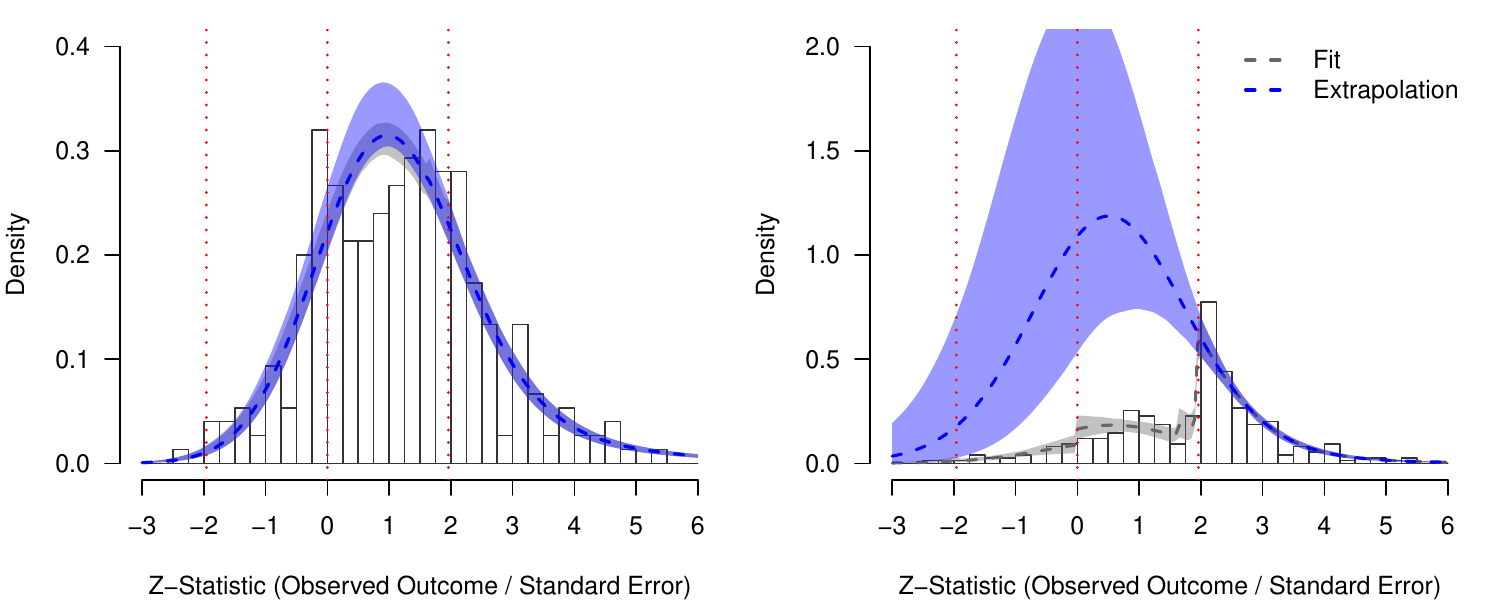}
    \caption{Model Fit Assessment and Extrapolation of Robust Bayesian Meta-Analysis Results z-plots on Simulated Datasets}
    \textit{Note.} The left panel visualizes the $300$ simulated studies in the absence of publication bias. The right panel visualizes the $182$ studies that survive under publication bias. The bands correspond to point-wise 95\% posterior predictive intervals.
    \label{fig:bias_no_bias3}
\end{figure}

Figure~\ref{fig:bias_no_bias3} shows the extrapolated predictive distribution (blue) alongside the fitted posterior predictive distribution from RoBMA. When publication bias is absent (left), the fitted and extrapolated posterior predictive distributions coincide. When publication bias is present (right), the extrapolated distribution is much higher in the area corresponding to non-significant results (down-weighted by the weight-function in the estimated model). The gray band denotes point-wise 95\% posterior predictive intervals (the central 95\% quantile of the predictive density at each test statistic). Uncertainty is larger for the extrapolated distribution under publication bias due to uncertainty in the selection-weight parameters. Note also how the extrapolated distribution from the model estimated under the presence of publication bias (right panel) matches the shape of the actual histogram in the absence of publication bias (left panel).

\subsection*{Additional Statistics}

The comparison of the posterior predictive distribution and the extrapolated posterior predictive distribution offers a new perspective on the data and the degree of publication bias. We define two metrics to describe characteristics of this comparison: the expected discovery rate (EDR) and the expected number of missing studies ($\text{N}_\text{missing}$).

The expected discovery rate corresponds to the expected proportion of statistically significant results in the absence of publication bias \citep{bartovs2020z}. In the meta-analytic settings, it can be computed as the proportion of the normalized extrapolated posterior predictive distribution, $I \, \pi^*$, beyond the conventional two-sided threshold (typically $|z|>1.96$),
\begin{equation*}
    \text{EDR} = \text{Pr}(\text{z}^*_\text{pred} > 1.96 \, \text{or} \, \text{z}^*_\text{pred} < -1.96 \mid \text{y}, \text{se}).
\end{equation*}
The EDR provides a bias-corrected estimate of the discovery rate (i.e., the percentage of statistically significant results that were obtained before publication). This estimate is valuable because it limits the false discovery rate (i.e., the percentage of statistically significant results that are false positives). \citet{soric1989statistical} developed a simple formula to compute the maximum false discovery rate that is compatible with the discovery rate; \citet{schimmack2023estimating} refer to this estimate as the false discovery risk because the actual rate of false positives remains unknown.\footnote{The EDR estimate can also be used to specify a new alpha level to achieve a desired false positive risk. This information can complement the interpretation of effect size estimates when there is considerable heterogeneity that suggests some effect sizes are in the wrong direction, zero, or too small to matter.}

The expected number of missing studies corresponds to the mismatch between the posterior predictive mass and the extrapolated posterior mass. With the observed number of studies $N$ and the normalizing constant $I$, the expected number of missing studies equals
\begin{equation*}
    N_\text{missing} = \left(I^{-1} - 1\right) N .
\end{equation*}
For non-selection models, the normalizing constant equals $1$ (i.e., no mismatch between the distributions) and the models therefore cannot predict any missing studies. Because \texttt{RoBMA} constrains the relative publication weight for statistically non-significant results to be at most $1$, the expected number of missing studies may be positively biased (it is non-negative). Note that this model-based estimate of the expected number of missing studies differs from fail-safe N \citep{rosenthal1979file}, which asks how many unpublished null results would render the pooled effect size estimate non-significant.\footnote{The \texttt{RoBMA} function allows fixing the mean effect at $0$. Estimating the expected number of missing studies under this constraint returns the number of missing studies under the null hypothesis given the fitted publication bias model.}


\setcounter{table}{0}
\setcounter{figure}{0}
\setcounter{section}{0}
\renewcommand{\thetable}{B\arabic{table}}
\renewcommand{\thefigure}{B\arabic{figure}}
\renewcommand{\thesection}{Appendix B}

\section{Example: Effect of Framing on Decision Making}

As another example we analyze registered replication reports of the classic framing effect on decision making \citep{tversky1981framing} conducted as part of the Many Labs 2 project \citep{klein2018many}. The dataset includes 55 estimates that, when pooled via a random-effects model, result in a pooled effect size estimate of $g=0.44$, 95\% CI $[0.38, 0.50]$.

\begin{figure}
    \centering
    \includegraphics[width=1\linewidth]{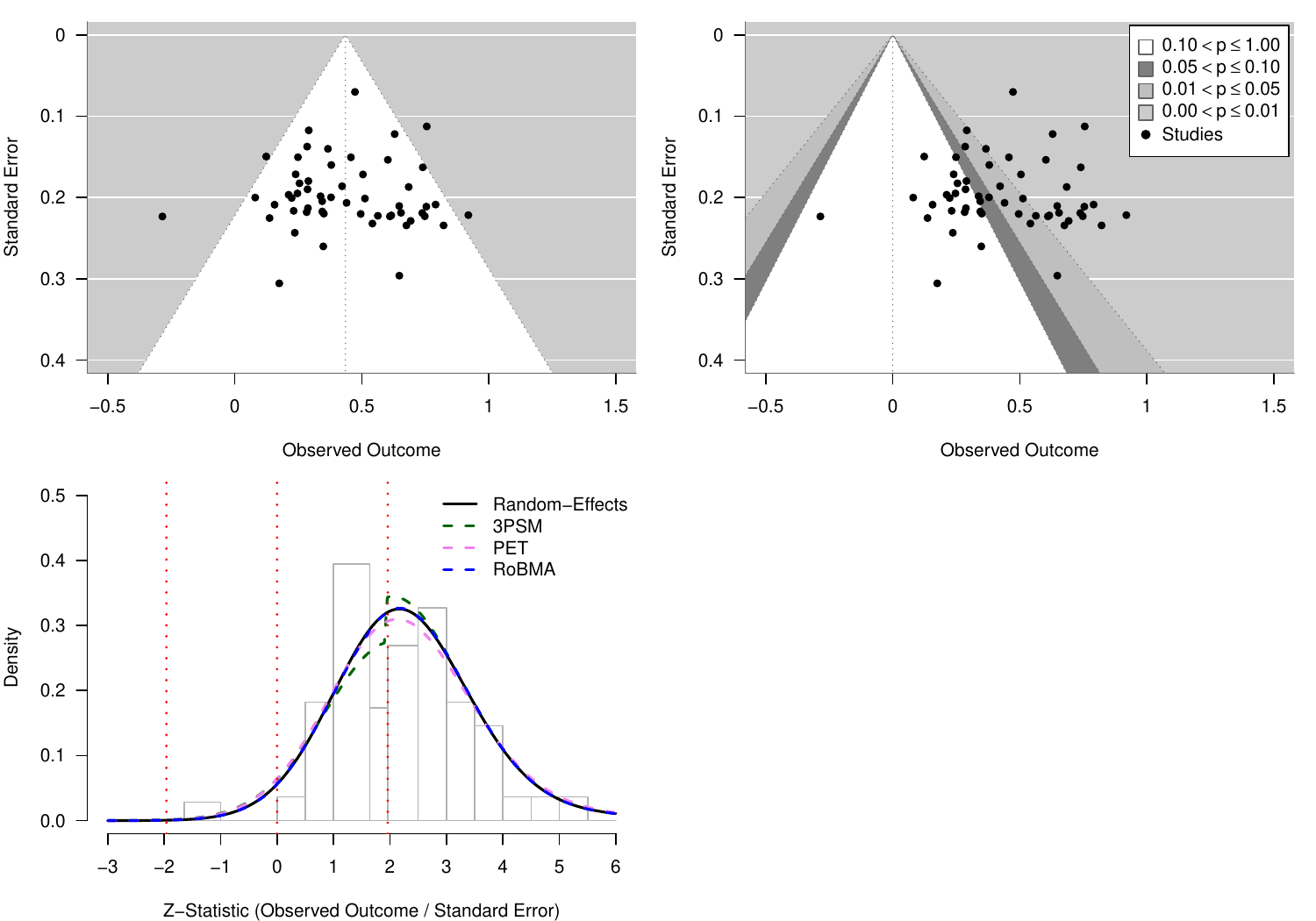}
    \caption{Funnel Plot, Contour Enhanced Funnel Plot, and z-plot of Registered Replication Reports of the Effect of Framing on Decision Making}
    \label{fig:example3}
\end{figure}

Neither the funnel plot (top panel of Figure~\ref{fig:example3}) nor the z-plot (bottom panel of Figure~\ref{fig:example3}) shows signs of publication bias. The model fits of random-effects, PET, 3PSM, and RoBMA are essentially identical and approximate the observed $z$-statistics well. Since any of these models would be appropriate for these data, the most parsimonious choice would be interpreting the random-effects model \citep{jefferys1992ockham}.

This conclusion aligns with the RoBMA result of moderate evidence against publication bias, BF\textsubscript{bias} $=0.21$. This result highlights how Bayesian model averaging penalizes unnecessary complexity (e.g., publication bias-adjusted models) when a parsimonious model explains the data equally well. Accordingly, the model-averaged effect is $g=0.42$, 95\% CI $[0.35, 0.49]$, with small between-study heterogeneity, $\tau=0.08$, 95\% CI $[0.00, 0.18]$, closely matching the unadjusted random-effects estimate.

The Many Labs 2 dataset includes 27 additional registered replication reports whose z-plots are available at \url{https://osf.io/pm45q/files/osfstorage}. Only three z-plots indicate a misfit of the random-effects model, accompanied by moderate evidence of publication bias. These false-positive results are unfortunately inevitable in any statistical analysis \citep{sterne2011recommendations}.

\clearpage

\bibliographystyle{biometrika}
\bibliography{bib/all.bib, bib/software, bib/unpublished} 

\end{document}